\documentclass[conference]{IEEEtran}

\usepackage[table]{xcolor}
\usepackage[hyphens]{url}
\usepackage{morefloats}
\usepackage{tocloft}
\usepackage[export]{adjustbox}
\usepackage{wrapfig}
\usepackage{lipsum}
\usepackage{soul}
\usepackage{breakcites}
\usepackage{varioref}
\usepackage{amsfonts}
\usepackage{amssymb}
\usepackage{pifont}
\usepackage{mathtools}
\usepackage{oplotsymbl}
\usepackage{xifthen}
\usepackage{xhfill}
\usepackage{geometry}
\usepackage[graphicx]{realboxes}
\usepackage{booktabs}
\usepackage{makecell}
\usepackage{multicol}
\usepackage{multirow}
\usepackage{colortbl}
\usepackage{varwidth}
\usepackage{setspace}
\usepackage{tabularx}
\usepackage{longtable}
\usepackage{arydshln}
\usepackage{pdfpages}
\usepackage[lined]{algorithm2e} \usepackage{semantic}
\usepackage{lastpage}
\usepackage{ragged2e}
\usepackage{hyphenat}
\usepackage[most]{tcolorbox}
\usetikzlibrary{backgrounds}

\definecolor{palette15}{HTML}{333333}
\definecolor{palette25}{HTML}{666666}
\definecolor{palette35}{HTML}{999999}
\definecolor{palette45}{HTML}{cccccc}
\definecolor{palette55}{HTML}{eeeeee}

\colorlet{text}{palette15}
\colorlet{titles}{palette25}
\colorlet{subtitles}{palette35}
\colorlet{subsubtitles}{palette45}

\usepackage{framed}

\newif\ifxetexorluatex
\ifxetex
  \xetexorluatextrue
\else
  \ifluatex
    \xetexorluatextrue
  \else
    \xetexorluatexfalse
  \fi
\fi
\ifxetexorluatex \usepackage{fontspec}
  \usepackage{libertine} \newfontfamily\quotefont[Ligatures=TeX]{Linux Libertine O} \else
  \usepackage[utf8]{inputenc}
  \usepackage[T1]{fontenc}
  \usepackage{libertine} \newcommand*\quotefont{\fontfamily{LinuxLibertineT-LF}} \fi

\newcommand*\quotesize{60} \newcommand*{\openquote}
   {\tikz[remember picture,overlay,xshift=-4ex,yshift=-2.5ex]
   \node (OQ) {\quotefont\fontsize{\quotesize}{\quotesize}\selectfont``};\kern0pt}

\newcommand*{\closequote}[1]
  {\tikz[remember picture,overlay,xshift=4ex,yshift={#1}]
   \node (CQ) {\quotefont\fontsize{\quotesize}{\quotesize}\selectfont''};}

\colorlet{shadecolor}{white!98!black}

\newcommand*\shadedauthorformat{\emph} 

\newcommand*\authoralign[1]{\if#1l
    \def\authorfill{}\def\quotefill{\hfill}
  \else
    \if#1r
      \def\authorfill{\hfill}\def\quotefill{}
    \else
      \if#1c
        \gdef\authorfill{\hfill}\def\quotefill{\hfill}
      \else\typeout{Invalid option}
      \fi
    \fi
  \fi}

\newcommand{\sidenote}[1]{\todo[size=\scriptsize,color=black!20!blue!10!white,bordercolor=black!20]{#1}}
\newcommand{\sn}[2][]{{\ifthenelse{\isempty{#1}}{\sidenote{#2}}{\sidenote{{\bf #1:} #2}}}}
\newcommand{\csn}[3][]{{\setulcolor{red}\ul{#3}}{\ifthenelse{\isempty{#1}}{\sidenote{#2}}{\sidenote{{\bf #1:} #2}}}}

\definecolor{ncl}{rgb}{0.86,0.06,0.18}
\definecolor{ati}{rgb}{0,0,0}
\definecolor{uow}{rgb}{0.1,0.36,0.68}
\definecolor{uos}{rgb}{0.78,0.65,0}

\newcommand{\hr}{\hbox to\headwidth{\color{palette55}\leaders\hrule height \headrulewidth\hfill}}
\newcommand{\uld}[1]{\parbox[b][-1pt][l]{0pt}{\color{palette55}\leavevmode\makebox[\widthof{#1\,}]{\xleaders\hbox{.}\hfill\kern0pt}}#1}
\newcommand{\well}[2][]{\bgroup\small\noindent \renewcommand{\arraystretch}{1.5}
\begin{tabular*}{\textwidth}{m{0.00001\textwidth}m{0.013\textwidth}m{0.87\textwidth}p{0.001\textwidth}}& \cellcolor{palette45}\color{white}\centering\rotatebox{90}{\textsc{\bfseries #1}}& {\color{gray}#2}& \end{tabular*}\egroup}

\hbadness = \tolerance
\colorlet{evenrows}{white!98!black}
\colorlet{oddrows}{palette55!90!blue!20!white}
\colorlet{headercolor}{palette55!95!blue}
\colorlet{headercoloralt}{palette55!95!green}

\newcommand{\rotation}{60}
\newcommand{\thr}[2][\rotation]{\rotatebox{#1}{\bfseries\color{titles}#2}}
\newcommand{\thh}[2][headercolor]{\cellcolor{#1}\bfseries\color{titles}#2}
\newcommand{\thc}[2][\centering]{\cellcolor{headercolor}#1#2}

\newcommand{\multilinecell}[4][evenrows]{\rowcolors{2}{#1}{#1}\renewcommand{\arraystretch}{1.1}\hspace{-.65em}\begin{tabular}[#2]{#3}#4\end{tabular}\hspace{-.65em}}

\newcommand{\cellwithnote}[3][.05\columnwidth]{\renewcommand{\arraystretch}{1}\hspace{-.65em}\begin{tabular*}{#1}{lr}#2 & \color{gray}\smaller\em#3\end{tabular*}}

\newcommand{\rb}{\circletfill}
\newcommand{\rh}{\circletfillhb}
\newcommand{\rt}{\circletfillha}
\newcommand{\rc}{\circlet}
\newcommand{\icon}[2][inline]{\ifthenelse{\equal{#1}{inline}}{\hspace{1.4em}\begin{tikzpicture}[remember picture,overlay]
  \node[] at (-0.9em,0.4em) { \includegraphics[width=1.3em]{#2} };
\end{tikzpicture}}{\begin{tikzpicture}[remember picture,overlay]
  \node[] at (#1) { \includegraphics[width=1.3em]{#2} };
\end{tikzpicture}}}

\newcommand{\clearcell}{\cellcolor{white}}

\newtcbox{\pill}[1][palette55]{nobeforeafter,tcbox raise base,arc=0.6em,outer arc=0.6em,top=0.1em,bottom=0.1em,left=0.2em,right=0.2em,leftrule=0mm,rightrule=0mm,toprule=0mm,bottomrule=0mm,boxsep=0.1em,colback=#1,colframe=#1,coltext=white,fontupper=\bfseries}

\newcolumntype{C}{>{\centering\arraybackslash\color{titles}\bfseries}m{0.17\textwidth}} \newcolumntype{D}{>{\raggedright\arraybackslash\color{text}}m{0.12\textwidth}} \newcolumntype{E}{>{\raggedright\arraybackslash\color{text}}m{0.41\textwidth}} \newcolumntype{M}{>{\centering\arraybackslash\color{text}}b{0.008\textwidth}} 

\newcolumntype{H}{>{\centering\arraybackslash\color{titles}\bfseries}m{8em}} \newcolumntype{I}{>{\raggedright\arraybackslash\color{titles}\bfseries}m{.4em}} \newcolumntype{N}{>{\arraybackslash\color{text}}m{0.6\textwidth}} 

\newcolumntype{Y}{>{\centering\arraybackslash\vspace{-1em}\color{green!70!black}$\blacktriangleright$}p{0.014\textwidth}} \newcolumntype{S}{>{\arraybackslash\vspace{-.5em}\color{text}}p{0.9\textwidth}} \newcolumntype{L}{>{\centering\arraybackslash\vspace{-.5em}\color{green!70!black}}p{0.014\textwidth}} \newcolumntype{Z}{>{\centering\arraybackslash\color{green!70!black}\openup .3em}p{0.05\textwidth}} \newcolumntype{R}{>{\arraybackslash\vspace{-1em}\color{text}}p{0.95\textwidth}}

\newcolumntype{T}{>{\raggedleft\arraybackslash\color{titles}\bfseries}p{0.22\textwidth}} \newcolumntype{F}{>{\raggedright\arraybackslash\color{text}}p{0.78\textwidth}} 

\newcolumntype{U}{>{\raggedleft\arraybackslash\color{titles}\bfseries}p{0.2\textwidth}} \newcolumntype{G}{>{\raggedright\arraybackslash\color{text}}p{0.8\textwidth}}

\begin{document}
\title{An Overview of Cyber Security and Privacy on the Electric Vehicle Charging Infrastructure}

\author{Roberto~Metere,
        Zoya~Pourmirza,
        Sara~Walker
        and~Myriam~Neaimeh\thanks{Roberto Metere is with the Department of Computer Science, University of York, York, UK. Large portion of this work has been done while Roberto was affiliated with both Newcastle University, Newcastle upon Tyne, UK, and The Alan Turing Institute, London, UK.}\thanks{Zoya Pourmirza, Sara Walker, and Myriam Neaimeh are with School of Engineering, Newcastle University, Newcastle upon Tyne, UK.}\thanks{Myriam Neaimeh is also with The Alan Turing Institute, London, UK.}}

\let\svthefootnote\thefootnote
\newcommand{\blankfootnote}[1]{%
  \let\thefootnote\relax\footnotetext{#1}%
  \let\thefootnote\svthefootnote%
}
\let\svfootnote\footnote
\renewcommand{\footnote}[2][?]{%
  \if\relax#1\relax%
    \blankfootnote{#2}%
  \else%
    \if?#1\svfootnote{#2}\else\svfootnote[#1]{#2}\fi%
  \fi
}
\newcommand{\inst}[1]{\textsuperscript{#1}}
\newcommand{\affiliation}[3]{\begin{tabular}{c}
  \inst{#1}{\em #2}\\#3
\end{tabular}}
\author{
\IEEEauthorblockN{Roberto Metere\inst{1,$\star$}, Zoya Pourmirza\inst{2}, Sara Walker\inst{2}, Myriam Neaimeh\inst{2,3}}
\IEEEauthorblockA{
\begin{tabular}{ccc}
  \affiliation{1}{University of York}{York, UK}
& \affiliation{2}{Newcastle University}{Newcastle upon Tyne, UK}
& \affiliation{3}{The Alan Turing Institute}{London, UK}
\end{tabular}
}
}

\markboth{Journal of \LaTeX\ Class Files,~Vol.~14, No.~8, August~2015}{Shell \MakeLowercase{\textit{et al.}}: Bare Demo of IEEEtran.cls for IEEE Journals}

\maketitle

\begin{abstract}
  Electric vehicles (EVs) are key to alleviate our dependency on fossil fuels.
The future smart grid is expected to be populated by millions of EVs equipped with high-demand batteries.
To avoid an overload of the (current) electricity grid, expensive upgrades are required.
Some of the upgrades can be averted if users of EVs participate to energy balancing mechanisms, for example through bidirectional EV charging.
As the proliferation of consumer Internet-connected devices increases, including EV smart charging stations, their security against cyber-attacks and the protection of private data become a growing concern.
We need to properly adapt and develop our current technology that must tackle the security challenges in the EV charging infrastructure, which go beyond the traditional technical applications in the domain of energy and transport networks.
Security must balance with other desirable qualities such as interoperability, crypto-agility and energy efficiency.
Evidence suggests a gap in the current awareness of cyber security in EV charging infrastructures.
This paper fills this gap by providing the most comprehensive to date overview of privacy and security challenges
To do so, we review communication protocols used in its ecosystem and provide a suggestion of security tools that might be used for future research.
 \end{abstract}


\footnote[]{\inst{$\star$} A significant part of this work was done while at Newcastle University and The Alan Turing Institute.}

\IEEEpeerreviewmaketitle

\section{Introduction}
\label{sec:intro}

The Climate Change Act of 2008 has proposed to reduce 100\% of the UK greenhouse gas emissions by 2050 (compared to 1990 levels)\cite{uk2008climate}.
A crucial part of the future smart cities that can contribute to this goal are zero emission vehicles, such as electric vehicles (EVs).
Indeed, EVs can decrease our dependency on fossil fuels and support smart\footnote{Throughout the paper, we use the attribute smart in its general meaning of being interconnected with other devices' networks, often to the Internet.} and flexible power systems.
The management of EVs should be able to prevent the electricity grid from overloading, thus minimizing expensive upgrades whose financial impact will be, in the end, suffered by consumers.
The panorama of EV chargers changed rapidly from having a few slow chargers in local (and relatively small) areas, that need not data links, to several charging stations across cities.
Such stations host a diversity of chargers, offering fast, smart charging with or without vehicle-to-grid capability (bidirectional energy transfer), and a multiplicity of services, where private information needs to be exchanged.
As other elements in a smart city\cite{laufs2020security}, EV charging infrastructures include smart functionality, and must ensure cyber security, data privacy, and interoperability.
Several studies outline a critical role for cyber security considerations for EV charging infrastructure, see Section~\ref{sec:related-work}.
Apart from a clear interest on the topic, the evidence reviewed here suggests that relatively little has been investigated about security in EV charging infrastructures.
This includes protocols between vehicles and chargers and between chargers and other entities, e.g., aggregators.
Additionally it is not clear what potential cyber attacks are more relevant or impactful to the EV charging infrastructure, nor what are the existing guidelines and standards that relate to providing a secure charging ecosystem.
This paper aims to fill all these gaps.

A security breach on communication devices in the EV ecosystem can have a significant impact on the Energy and Transport sectors\cite{zeller2011myth,whitehead2017ukraine}.
Securing the EV ecosystem is of utmost importance since Energy and Transport are usually national critical infrastructures, e.g., in the UK, they are two of the thirteen national infrastructure sectors recognised as critical.
The loss or compromise of critical infrastructures could impact on national security and on the availability, integrity or delivery of essential services that are crucial for a modern country to function and its people's daily life.

The EV ecosystem is populated by a considerable number of interconnected devices and entities that expose a large potential attack surface.
Attackers can be insiders, e.g., a legitimate employee illegally selling corporate data to third parties, or outsiders, e.g., nation-state backed cyber groups.
As EV charging infrastructures are part of energy systems, many of cyber security and data privacy concerns related to energy systems apply\cite{ebrahimy2017cyber,cameron2018using}.
For example, the smart EV chargers that connect vehicles to the power grid make use of an additional data connection to exchange information and control commands between various entities in the ecosystem; this makes them potential targets for cyber security attacks.

For the future smart city to work properly, confidential data needs to be exchanged and its privacy must be preserved.
Data privacy issues relate to the messages that need to be exchanged across the whole EV charging ecosystem to enable for a plurality of services to consumers and the power grid (ancillary services).
Such messages embed confidential information and, thus, need not only to be protected from leakage by latest cryptographic best practices, but also to be treated according to data privacy policies, e.g., General Data Protection Regulation (GDPR).
A subset of the data manipulated is users' data, such as their identity, their vehicle and their own location, bank account information, to name a few.
Similar private information are used in several payment systems already, where secure communications are usually sufficient measures: conversely in the EV charging infrastructure, devices holding private information, e.g., chargers and vehicles, are left unattended and exposed to the public for very long periods of time.
Additionally, the high mobility nature of EVs sees EV attaching to charging stations that need to be trusted.
Trusting the infrastructure by all participants is fundamental: a distrusting user may decide not to participate in managed charging initiatives, with a consequent potential disruption to the services impacting other users.
Ensuring both cyber security and data privacy is key to gain this trust.

 \subsection[The EV ecosystem]{The electric vehicle ecosystem}
\label{sec:ev-overview}

\begin{figure}[!ht]
  \begin{center}
\includegraphics[width=\columnwidth]{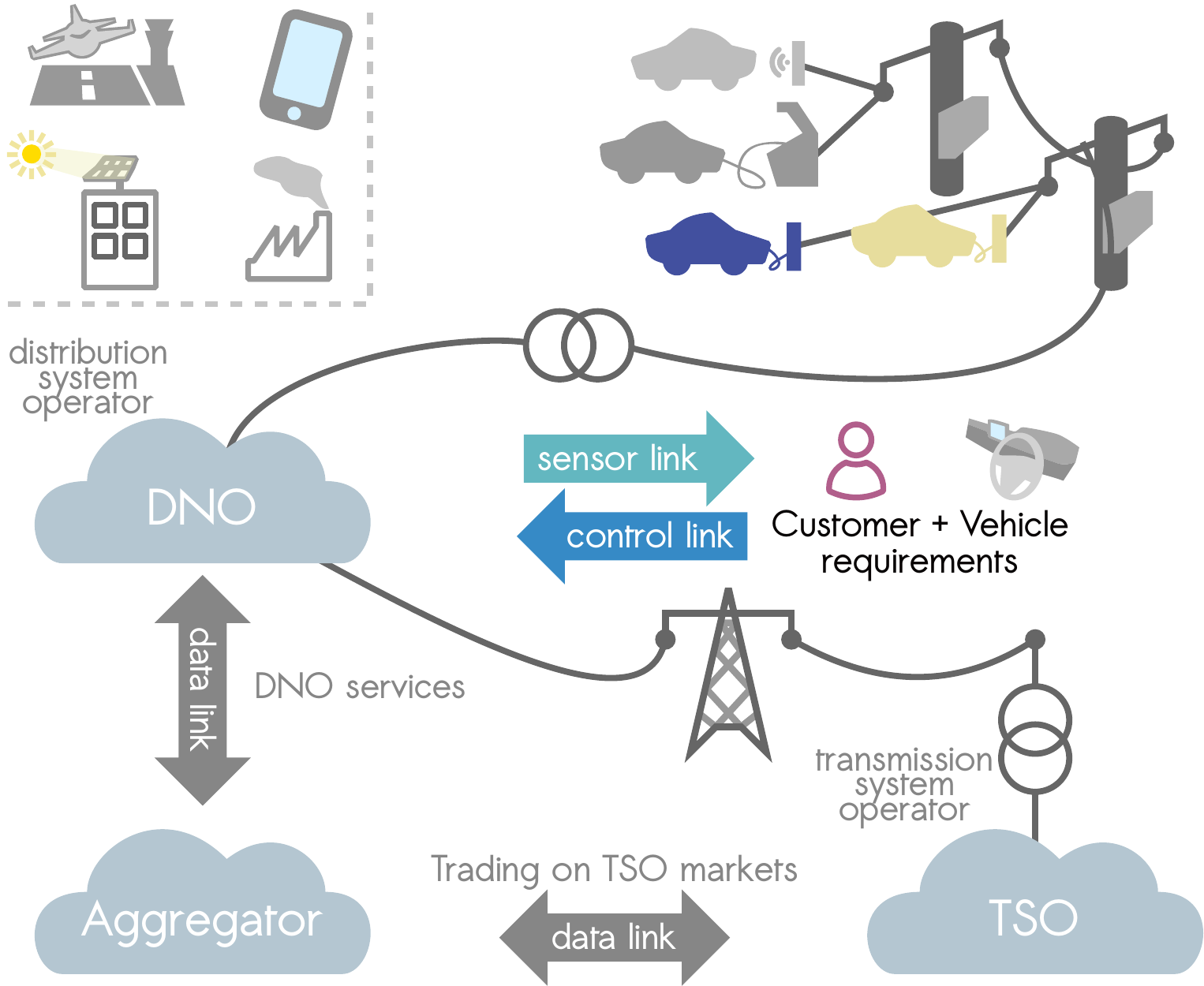}
  \end{center}
  \caption{An overview of a smart EV charging infrastructure.}
  \label{fig:full-panorama-ev-charging}
\end{figure}

The EV ecosystem is integrated into electricity systems and populated by many actors, as illustrated in Fig.~\ref{fig:full-panorama-ev-charging}: EVs (and their drivers), manufacturers, charging point operators, aggregators, distribution operators, transmission operators.

Charger points communicate with different entities in the EV ecosystem including vehicles, charging stations, charging station operators, grid network operators, among others.
For example, to charge an EV, coordination and communication is required between the vehicle, a charger in the charging station, whose system is managed by an operator that can aggregate multiple chargers across multiple charging stations to offer ancillary services; so, the aggregator will also communicate with distribution operators and, possibly, with transmission operators.

\subsection[The EV charging infrastructure]{The electric vehicle charging infrastructure}
\label{sec:evci-overview}

The EV charging infrastructure is a significant part of the EV ecosystem.
Communication protocols are used to facilitate the charging management strategies.
The EV charging infrastructure enables the connection between EVs and the electricity grid via charger points.
When relevant, e.g., when vehicle-to-grid is enabled, the EV batteries can buy and sell energy; thus, users of EVs are actively (but indirectly) exposed to the electricity market.
EVs can also offer services to the building or home (behind-the-meter services), to the neighbourhood or local electricity networks (distribution system operator services) and to the whole system or region (transmission system operator services)\cite{andersen2019parker,neaimeh2020mind}.

Communication protocols specify the rules to follow for exchanging messages between entities.
Protocols can be designed for only part of the EV charging infrastructure, e.g., CHAdeMO specifies the communication between the EV and the charger, and they may or may not be interoperable with other protocols, for specific services that need to involve multiple entities.
This may trivially hinder security if information must flow unprotected in some network communication links, i.e., CHAdeMO does not protect information.
Additionally, interoperability allows for a diversity of entities, e.g. a charger can be replaced with a charger from another manufacturer as long as it can operate with the required charging protocols.
These protocols are either proprietary, usually developed and sold by private groups, or open, which may be developed by an accredited standard organisation.
For open protocols, either their specifications or their implementation (code) -- or both -- are accessible at no (or minimal) cost.
Open protocols may ease interoperability, as developers of other protocols can access to their specifications; they can also play a role in customer's trust.

A lack of interoperability may hinder security and privacy from at least two points of view: first, composition of secure protocols is not automatically secure (see Section~\ref{sec:security-analysis-communication-protocols}), and, second, regulations may simply mandate the use of specific protocols that may result non interoperable.
The former is well known in the security community; the latter becomes non-trivial in the EV charging infrastructure, as it is regulated by different (governmental) entities that do not usually work together, e.g., the Energy department and the Transportation department.

 \section[Vulnerabilities on the EV charging infrastructure]{Vulnerabilities in the electric vehicle charging infrastructure}
\label{sec:evci-cybersecurity}

The EV charging infrastructure populates the legacy grid (power grid) with new entities, services, and adds a communication layer whose entities can be connected to the Internet.
As such, it increases the attack surface (especially from remote attacks) introducing new potential vulnerabilities allowing for a series of cyber-attacks.
Such attacks could be perpetrated through communication channels (as for in the Internet) and aim to provoke disruption to the power grid and its services, or more in general to the Energy and Transport sectors.
A synchronised and well-orchestrated attack may exploit multiple vulnerabilities: this is particularly worrying as the physical dimension of energy systems is prone to cascading effect in case of (targeted) failures\cite{palleti2021cascading}.
Cyber-attacks can be perpetrated in various ways, some of which go beyond technical vulnerabilities.
In this section we discuss cyber vulnerabilities in EV charging infrastructures and security standards that could help secure this infrastructure.

Recent studies\cite{metere2021securing,pourmirza2021electric} identified the key cyber security challenges of the complex EV charging ecosystem.
They are: physical limitations of devices and communication channels; human factor, heterogeneity, scale, and ad-hoc nature of threats; authentication and identity management; authorisation and access control; and updating, responsibility, and accountability.
Vulnerabilities in the above challenges potentially put systems at risks of cyber attacks.

To identify the risks of a system, a traditional approach is to break down the problem in several categories of desired security goals such as the well-established CIA triad: confidentiality, integrity and availability.
Confidentiality is to make sure that some data is only accessible by authorised users.
Integrity is to enforce trust in the information that is flowing, and ensure modification can only be done by authorised agents.
Availability is to make sure the information is available in a timely manner and responds to the queries of the user within an expected time frame (e.g., Denial of Service (DoS) attack).

In Table~\ref{tbl:attack-type}, we identified possible attacks according to the security goals of confidentiality, integrity and availability, and the impact of attacks in terms of physical impact, cyber impact or social impact.
\begin{table*}[!ht]
  \caption{Possible security attacks classified according with the CIA triad and impact.\\\underline{Legend}. Security: C -- confidentiality, I -- integrity, A -- availability; Impact: S -- Social, C -- cyber, P -- physical.}
  \label{tbl:attack-type}
  \smaller\sffamily
  \renewcommand{\arraystretch}{1.}
  \setlength\arrayrulewidth{2pt}
  \arrayrulecolor{white}
  \rowcolors{2}{oddrows}{evenrows}
  \begin{tabular}{C|D|M|M|M|M|M|M|E}
    & 
    & \thr[0]{C}
    & \thr[0]{I}
    & \thr[0]{A}
    & \thr[0]{S}
    & \thr[0]{C}
    & \thr[0]{P}
&  \\
      \thh{Attack type}
    & \thh{Specific target (optional)}
    & \multicolumn{3}{c|}{\thh{Security}}
    & \multicolumn{3}{c|}{\thh{Impact}}
    & \thh{Example} \\
    \hline
    
    \thc{Denial of Service (DoS)} & 
      &       &       & $\rb$ & $\rb$ & $\rb$ &       & Communication broken either at aggregator level or broader \\ \hline
    \thc{Bruteforcing} & EV
      & $\rb$ &       &       & $\rb$ & $\rb$ & $\rb$ & Guess passwords used in the charger's network \\ \hline
    \thc{Delay attack} & 
      &       &       & $\rb$ &       & $\rb$ & $\rb$ & Power requests at incorrect timing may cause breakdowns \\ \hline
    \thc{Replay attack} & 
      &       & $\rb$ &       &       & $\rb$ & $\rb$ & Incorrectly replaying power requests may cause breakdowns \\ \hline
    \thc{Snooping} & 
      & $\rb$ &       &       & $\rb$ &       & $\rb$ & An attacker may link messages to the same user to stalk, track, profile, or learn life habits (among others) \\ \hline
    \thc{Sybil attack} & Aggregator
      &       & $\rb$ &       & $\rb$ &       &       & Copy ID tokens to multiply energy charge for free \\ \hline
    \thc{Impersonation} & EV, aggregator
      &       & $\rb$ &       &       & $\rb$ &       & Steal energy from either directions \\ \hline
    \thc{Cloning} & EV owner
      & $\rb$ & $\rb$ &       & $\rb$ & $\rb$ &       & If an RFID tag is subject to duplication, an intruder could identify themself as the tag owner \\ \hline
    \thc{Man-in-the-middle} & 
      & $\rb$ & $\rb$ &       &       & $\rb$ & $\rb$ & Tamper with messages so that charging controls can be sent in reverse \\ \hline
    \thc{Repudiation} & EV
      &       &       & $\rb$ &       & $\rb$ &       & Denial to charge a legitimate EV \\ \hline
    \thc{EV misbehave} & Aggregator/billing
      &       & $\rb$ &       &       & $\rb$ & $\rb$ & Adding nois can lead to wrong information yielding wrong decisions \\ \hline
    \thc{Misinformation} & EV owner
      &       & $\rb$ &       & $\rb$ &       & $\rb$ & EV owners fed with fake information (incentive to charge at peak times) can change their behaviour in a way that corresponds to an attack to the grid \\ \hline
    \thc{Load-changing attack}  & EV, aggregator
      &       & $\rb$ & $\rb$ & $\rb$ & $\rb$ & $\rb$ & Synchronise charging or discharging operations to provoke unbalance in the power grid \\ \hline
  \end{tabular}
\end{table*}
Detecting cyber-threats against EV ecosystems is a complicated task, due to the large spread of the attack surface, and evolving EV charging infrastructures.
The attack surface is enlarged even more if we consider coexisting networks that partially manipulate the same data\cite{laufs2020security}. 
We can study the cyber-attack against such systems from various perspectives.
For example from an administrative perspective, a representative infrastructure and associated cyber-attacks against the power grid can be simulated in a virtual environment to evaluate and train operators to identify, detect and respond in a timely manner.
From an end-user perspective it is essential to educate the users to a variety of risks, which remain to be identified (e.g., connecting a vehicle to open wireless hotspots in the vicinity of charging stations).
From a technical perspective, EVs and the grid must encompass a range of traditional measures and protection devices such as firewalls, intrusion detection systems, disabling default credentials, encryption of communications, among others.

In this context, we emphasise that batteries in EVs can quickly transfer an amount of energy comparable to what several households averagely transfer in a much longer period of time.
As such, some attacks, as load-changing attacks\cite{arnaboldi2020modelling} (see Table~\ref{tbl:attack-type}), are far more impactful if they target EVs rather than traditional residential load.

\section{Standards for EV infrastructure}
As the proliferation of consumer Internet-connected devices increases, as traditionally offline systems have now become online.
To ascertain that enough measures have been taken to provide security, a number of standards have been developed in different contexts.
We briefly discuss security considerations of some of them, whose context relates to the EV charging infrastructure.

{\bf ENCS and ElaadNL\cite{encs2019ev}}.
The European Network for Cyber Security (ENCS) in collaboration with ElaadNL published several technical documents offering insights and considerations to develop secure charging infrastructure.
These include a security test plan for EV charging stations and a proposed security architecture for charging infrastructure.
Among their documents, they offer guidelines for a future-proof design for EV charging devices, and propose the designs of a peer-to-peer public-key infrastructure (PKI)\cite{elaadnl2018exploring} compatible and coexisting with the PKI proposed by ISO 15118.
A brief description of public-key infrastructures is provided by Metere et al.\cite{metere2021securing}.

{\bf BSI PAS 1878 and 1879}.
Both PAS 1878\cite{bsi2021pas1878} and 1879\cite{bsi2021pas1879} are dedicated to energy smart appliances, including the EV charging infrastructure and the V2G technology.
BSI PAS 1878 focuses on system functionality and architecture, while BSI PAS 1879 provides a code of practice on the demand side response operations.
They propose an end-to-end security framework that secures assets such as energy smart appliances (ESA) (e.g., EV chargers) and consumer energy managers (CEM), verifying actors such as demand-side response providers (DSRSP), and securing communications between ESA and CEM, or between CEM and DSRSP.
Without explaining all details, we suggest BSI PAS 1879\cite[Table A1]{bsi2021pas1879} for a quick reference of their security specifications.

{\bf National Institute of Standards and Technology (NIST)}.
The Information Technology Laboratory (ITL) is one of the six laboratories at NIST, which is responsible to develop reference data, tests, methods, proof of concept implementations, and technical analyses to improve the high-quality, independent, and unbiased research and data.
NIST standards are in-line and consistent with standards published by the International Organization for Standardization (ISO) and International Electrotechnical Commission (IEC).
For a brief overview on the cyber security framework from NIST, we refer to Metere et al.\cite{metere2021securing}.

{\bf Cyber Essential and ISO 27001}.
In 2014 the UK government launched an assurance scheme called Cyber Essential (CE)\cite{ncsc2018cyber} that is the minimum certification required for a government supplier handling personal information.
CE identified five technical cyber security controls that could be implemented by organizations to provide a baseline for cyber security.
In 2005 an international standard for information security was introduced, ISO/IEC 27001\cite{isoiec27001:2013}, that specifies best practices for information security management systems.
ISO 27001 does not have specific requirements for compliance similar to the CE and does not apply to small and medium-sized enterprises (SMEs) working in the UK.
Pourmirza and Walker\cite{pourmirza2021electric} provide further comparison between CE and ISO 27001.

We have two final remarks.
The first remark is that entities in the EV charging infrastructure will also be connected to IoT devices, e.g., EVs use sensors and can be connected to (home) networks with smart appliances.
As such, they are also expected to be aware of threats that may come from those additional networks, and that can be regulated by different policies, e.g., the cyber security for Internet of Things (IoT) published by the European Telecommunications Standards Institute (ETSI)\cite{etsi2019cyber,etsi2020cyber}.
The second remark is about the ongoing activity to establish regulation.
The publication of the European Cyber Security Act has initiated the work on the Network Code on energy-specific cyber security\footnote{\url{https://ec.europa.eu/energy/sites/ener/files/sgtf_eg2_2nd_interim_report_final.pdf}} which is currently under development.
Network Code has legal status within the EU; as previously part of the EU, the UK has implemented antecedent related documents.
 \section[Privacy challenges]{Privacy challenges}
\label{sec:evci-privacy}

The EV ecosystem, as part of the smart grid, is currently offering an increasing number of services; consequently, the amount of sensible data required to enable such services increases too.
This is particularly remarked by the adoption emerging technologies, e.g. vehicle-to-grid, that promote consumers (who merely consume energy) to {\em prosumers} (who can provide energy to other consumers).

Private information are exchanged between various entities in the EV ecosystem with the aim of providing accuracy and reliability of services.
Related communications must preserve privacy\cite{khan2021privacy,farao2021p4g2go} and develop into the already existing (unrelated) services offered to end-users, e.g., map directions and remote control.
Communications typical of the broader context of sustainable resilient cities\cite{ferrag2018systematic} partly show similar privacy concerns; though we can be more detailed as our context is narrower.

Mobile users are familiar with providing private data (i.e., GPS location) in return of an improved personalised service (e.g., traffic-aware driving directions, locations of nearby restaurants, among others).
Especially after recent regulations on privacy, e.g. GDPR, it becomes more and more common for users to be able to opt out from sharing (some of) their data.
A service that respects users' preferences may work with limited functionality or not work at all.

Similarly to mobile services, drivers of electric vehicles can participate to (smart) charging services to reduce the cost of their movements.
To enjoy an optimal service, users would put their trust to operators to control their charging operations and log their driving patterns (i.e., time of arrival, departure).
Apart from the expected personally identifiable information that can be involved, other usage data that does not immediately identify as personal can still be collected and further analysed to extract life patterns and habits, which are of private concerns.
If protecting data communication is thus a key requirement for a secure EV ecosystem, so is security of storages, databases and other memories that store private information temporarily (e.g., cache memory).

\begin{figure}[htbp]
  \begin{center}
    \includegraphics[width=\columnwidth]{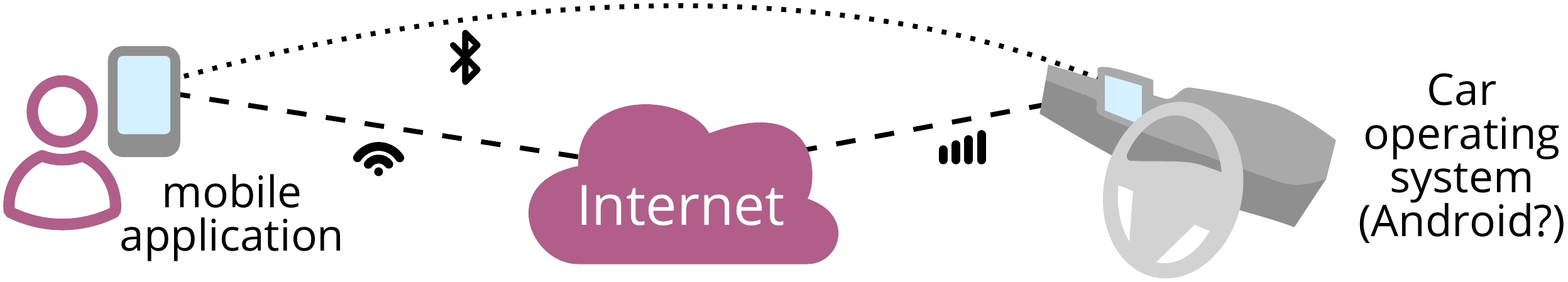}
  \end{center}
  \caption{Scheme of communication between the vehicle and its driver.}
  \label{fig:car-remote-control}
\end{figure}

The impact and importance of privacy is augmented in those cases where a vehicle communicates with a mobile device, and both are often connected to the Internet, see Fig.~\ref{fig:car-remote-control}.
Such communication is necessary to offer monitoring and (remote) control functionality, e.g., locking doors, climate, monitoring, park location, assistance, or charging preferences.
Modern vehicles can be constantly connected to the Internet; the driver of an EV can share their schedule with the charging operator through their smart phones.
This would be done to allow the charging operator to set a charging profile adequate and respectful to upcoming trips of the driver.

Not all manufacturers offer the same services, and thus they do not require the same pieces of user's personal data.
Table~\ref{tbl:mobile-permissions-android} summarises the permissions that are required from Android applications.
\begin{table}[!ht]
  \caption{Summary of permissions of some Android applications to monitor and remote control cars (as of 2020).}
  \label{tbl:mobile-permissions-android}
  \renewcommand{\rotation}{90}
  \centering
  \small\sffamily
  \renewcommand{\arraystretch}{1.}
  \setlength\arrayrulewidth{2pt}
  \arrayrulecolor{white}
  \rowcolors{2}{oddrows}{evenrows}
  \begin{tabular}{H|IIIIIIIIII}
  & \thr{Nissan} & \thr{Peugeot} & \thr{Vauxhall} & \thr{Tesla} & \thr{Toyota} & \thr{BMW} & \thr{Audi,Mini} & \thr{Renault} & \thr{Jaguar} & \thr{VW} \\ \toprule
  \thh{DeviceID \& call info} & $\rb$ & $\rb$ & $\rb$ &       & $\rb$ & $\rb$ & $\rb$ &        &       & $\rb$ \\ \hline
  \thh{Phone accounts       } &       &       &       &       &       &       & $\rb$ &        &       &       \\ \hline
  \thh{Phone status         } & $\rb$ & $\rb$ & $\rb$ &       & $\rb$ & $\rb$ & $\rb$ &        &       & $\rb$ \\ \hline
  \thh{Location             } & $\rb$ & $\rb$ & $\rb$ & $\rb$ & $\rb$ & $\rb$ & $\rb$ & $\rb$  & $\rb$ & $\rb$ \\ \hline
  \thh{Calendar             } &       & $\rb$ & $\rb$ & $\rb$ & $\rb$ & $\rb$ & $\rb$ &        &       &       \\ \hline
  \thh{Photos/ Media/Files  } & $\rb$ & $\rb$ & $\rb$ & $\rb$ & $\rb$ & $\rb$ & $\rb$ & $\rb$  & $\rb$ & $\rb$ \\ \hline
  \thh{Contacts             } & $\rb$ &       &       & $\rb$ & $\rb$ &       & $\rb$ &        &       &       \\ \hline
  \thh{Camera/ Microphone   } & $\rb$ & $\rb$ & $\rb$ &       & $\rb$ & $\rb$ & $\rb$ & $\rb$  &       &       \\ \hline
  \thh{Full network access  } & $\rb$ & $\rb$ & $\rb$ & $\rb$ & $\rb$ & $\rb$ & $\rb$ & $\rb$  & $\rb$ & $\rb$ \\ \hline
  \thh{Prevent sleep        } & $\rb$ & $\rb$ & $\rb$ & $\rb$ & $\rb$ & $\rb$ & $\rb$ & $\rb$  & $\rb$ & $\rb$ \\ \hline
  \thh{Device \& app history} &       & $\rb$ & $\rb$ &       &       &       &       &        &       &       \\
  \bottomrule
  \end{tabular}\end{table}
As an example of such services, most mobile apps by manufacturers (e.g.,Tesla, Audi, Toyota and others) require read-and-write permissions to the user's calendar.
Certainly, one must be cautious when accepting permissions in mobile applications; though, granting permissions does not mean that all permitted functions will be used.
For example, when an user grants an application to access their calendar, the application will unlikely retrieve all their past and future events from the calendar, or share them to some third party.
Yet, the granted permissions might allow to do so.
This caveat is linked to the relevance of offering open source implementations, where an (experienced) user can practically inspect the source code, see what information are stored, and where do they migrate to (e.g., sent to external servers).
We strongly recommend servers that collect user's data to {\em anonymise} and {\em aggregate} such data to protect their privacy.

Misuses of user's data are (obviously) negatively perceived by users; as a direct implication, they might resist in participating in necessary initiatives, e.g., smart charging.
Those who distrust ``smart'' system would not easily adopt to them.
Learning from the past, we refer to when smart meters were rolled-out in several countries.
Their deployment overlooked non-technical factors, such as social and ethical considerations.
This had a considerable impact on the (immediate) success of smart meter adoption and, as a consequence, could have diminished the trust to future smart energy system.

{\bf Scheduling of charging sessions}.
As electric vehicles can demand for a significant amount of energy, it is fundamental to schedule charging operations to avoid unnecessary situations of over-demand on peak times.
Important in this context (but not only) is the emerging technology of vehicle-to-grid (V2G) that enables for bidirectional energy transfer in EVs.
Thus it also enables aggregators to provide ancillary services to aid the reliability of the power grid, such as frequency regulation, voltage regulation, power-loss replacements or other services.
Those services, along with the potential access by private users (e.g., owners of the electric vehicles) to the energy market, need to be regulated and controlled.
Many algorithms to enact those services or to obtain optimal performances toward some desirable goal (i.e., financial, resilience, and others) have been proposed in the literature, see Table~\ref{tbl:services-data-scheme}. 

From the perspective of cybersecurity, bidirectional charging requires an additional communication layer where new sensible data flows and needs to be protected.
Failing to protect such data will permit confidential information to migrate elsewhere, e.g., through coexisting communication networks in vehicles\cite{abualola2021v2v}.
Thus, it is important to consider what are the data required by those algorithms to run, and determining what part of data are sensible relating to privacy.
The most relevant parameters to privacy we consider are:
{\em capacity} of the EV battery,
{\em state of charge} of the EV battery,
{\em charging rate} of the EV battery (often the output of the algorithm -- negative rate would mean discharging),
{\em arrival/departure time} when the EV battery is attached to the charger, and when it plans to leave,
{\em network topology} of aggregators and chargers,
{\em aggregator income} or {\em cost} relating to the revenue when trading in the energy market,
{\em price offered to EVs} when the energy market is mediated by aggregators which can establish custom prices,
{\em system load} the demand of a grid node (e.g., distribution), and
{\em trip distance} or {\em time} the distance planned, or the maximum allowed by a full tank and full battery, and the time required to cover the distance.
We summarised scheduling algorithms proposed in the literature, especially those relating to V2G technology, in Table~\ref{tbl:services-data-scheme}.
Those algorithms are mainly focused on two aspects: optimisation economical revenues, by either participating in ancillary services or optimising the usage of hybrid engines and trips; and involving V2G into ancillary services, that are offered by regulating the charging rate of electric vehicles.

\begin{table*}[htbp]
  \renewcommand{\rotation}{74}
  \centering
  \small\sffamily
  \renewcommand{\arraystretch}{1.}
  \setlength\arrayrulewidth{2pt}
  \arrayrulecolor{white}
  \rowcolors{2}{oddrows}{evenrows}
  \caption{Summary of data used by proposed algorithms. The symbol $\rh$ is used when the value is used but as average, approximate or a relatable value is required, otherwise $\rb$ when it is directly required; if the value is not required, we use $\rc$.}
  \label{tbl:services-data-scheme}
  \begin{tabular}{cc|l|l|l|l|l|l|l|l|l|l|l}
  \clearcell & & \thr{capacity} & \thr{state of charge} & \thr{charging rate} & \thr{arrival/departure time} & \thr{network topology} & \thr{aggregator income/cost} & \thr{price offered to EVs} & \thr{system load} & \thr{trip distance/time} \\ \toprule
  \clearcell & \thh{Han      et al.\cite{han2010development}                     } & $\rb$ & $\rb$ & $\rb$ & $\rc$ & $\rc$ & $\rc$ & $\rc$ & $\rc$ & $\rc$ \\
  \clearcell & \thh{Sortomme et al.\cite{sortomme2011optimal}                    } & $\rb$ & $\rb$ & $\rb$ & $\rc$ & $\rc$ & $\rc$ & $\rb$ & $\rc$ & $\rc$ \\
  \clearcell & \thh{Lunz     et al.\cite{lunz2011optimizing}                     } & $\rc$ & $\rb$ & $\rc$ & $\rc$ & $\rc$ & $\rc$ & $\rc$ & $\rc$ & $\rb$ \\
  \clearcell & \thh{Shi      et al.\cite{shi2011real}                            } & $\rb$ & $\rb$ & $\rb$ & $\rb$ & $\rc$ & $\rc$ & $\rc$ & $\rc$ & $\rc$ \\
  \clearcell & \thh{Sortomme et al.\cite{sortomme2012intelligent}                } & $\rb$ & $\rb$ & $\rb$ & $\rc$ & $\rc$ & $\rb$ & $\rc$ & $\rb$ & $\rc$ \\
  \clearcell & \thh{Mal      et al.\cite{mal2013electric}                        } & $\rb$ & $\rb$ & $\rb$ & $\rb$ & $\rc$ & $\rc$ & $\rc$ & $\rc$ & $\rc$ \\
  \clearcell & \thh{Turker   et al.\cite{turker2013housing}                      } & $\rb$ & $\rb$ & $\rb$ & $\rb$ & $\rc$ & $\rc$ & $\rc$ & $\rc$ & $\rc$ \\
  \multirow{-5}{*}{\clearcell\thr[90]{2010-2013\quad}} & \thh{Mohamed  et al.\cite{mohamed2013real}                        } & $\rb$ & $\rb$ & $\rb$ & $\rh$ & $\rc$ & $\rc$ & $\rc$ & $\rc$ & $\rc$ \\ \arrayrulecolor{headercolor}  \hline \arrayrulecolor{white} \hline
  \clearcell & \thh{Ansari   et al.\cite{ansari2014optimal}                      } & $\rb$ & $\rb$ & $\rb$ & $\rc$ & $\rc$ & $\rb$ & $\rc$ & $\rc$ & $\rc$ \\
  \clearcell & \thh{Lin      et al.\cite{lin2014optimal}                         } & $\rc$ & $\rb$ & $\rb$ & $\rc$ & $\rc$ & $\rc$ & $\rc$ & $\rc$ & $\rc$ \\
  \clearcell & \thh{Ansari   et al.\cite{ansari2014coordinated}                  } & $\rb$ & $\rc$ & $\rb$ & $\rb$ & $\rc$ & $\rb$ & $\rc$ & $\rc$ & $\rc$ \\
  \clearcell & \thh{Kavousi  et al.\cite{kavousi2015reliability}                 } & $\rb$ & $\rb$ & $\rc$ & $\rc$ & $\rc$ & $\rc$ & $\rc$ & $\rc$ & $\rc$ \\
  \clearcell & \thh{Yu       et al.\cite{yu2015balancing}                        } & $\rc$ & $\rb$ & $\rc$ & $\rh$ & $\rb$ & $\rc$ & $\rc$ & $\rb$ & $\rc$ \\
  \multirow{-5}{*}{\clearcell\thr[90]{2014-2017\quad}}\clearcell & \thh{Lam      et al.\cite{lam2017coordinated}                     } & $\rb$ & $\rb$ & $\rb$ & $\rc$ & $\rb$ & $\rc$ & $\rc$ & $\rc$ & $\rb$ \\ \arrayrulecolor{headercolor}  \hline \arrayrulecolor{white} \hline
  \clearcell & \thh{Ko       et al.\cite{ko2018mobility}                         } & $\rc$ & $\rb$°& $\rb$ & $\rc$ & $\rb$ & $\rc$ & $\rc$ & $\rc$ & $\rc$ \\
  \clearcell & \thh{Turker   et al.\cite{turker2018optimal}                      } & $\rb$ & $\rb$ & $\rb$ & $\rb$ & $\rc$ & $\rc$ & $\rc$ & $\rc$ & $\rc$ \\
  \clearcell & \thh{Chen     et al.\cite{chen2018online}                         } & $\rb$ & $\rb$ & $\rb$ & $\rb$ & $\rc$ & $\rc$ & $\rc$ & $\rb$ & $\rc$ \\
  \clearcell & \thh{Li       et al.\cite{li2020boosting}                         } & $\rb$ & $\rb$ & $\rb$ & $\rb$ & $\rc$ & $\rc$ & $\rc$ & $\rc$ & $\rc$ \\
  \clearcell & \thh{Liang    et al.\cite{liang2020mobility}                         } & $\rc$ & $\rb$ & $\rc$ & $\rb$ & $\rb$ & $\rc$ & $\rh$ & $\rc$ & $\rc$ \\
  \clearcell & \thh{Jin      et al.\cite{jin2020optimal}                         } & $\rh$ & $\rb$ & $\rb$ & $\rb$ & $\rh$ & $\rc$ & $\rb$ & $\rh$ & $\rc$ \\
  \clearcell & \thh{Zhou     et al.\cite{zhou2020coordinated}                    } & $\rh$ & $\rb$ & $\rh$ & $\rb$ & $\rc$ & $\rc$ & $\rc$ & $\rb$ & $\rc$ \\
  \multirow{-5}{*}{\clearcell\thr[90]{2018-2021\quad}}& \thh{Kasani   et al.\cite{kasani2021optimal}                      } & $\rb$ & $\rb$ & $\rc$ & $\rb$ & $\rc$ & $\rc$ & $\rc$ & $\rb$ & $\rc$ \\ \arrayrulecolor{headercolor}  \hline
\end{tabular}\vskip 0.5em
  \hfill\scriptsize °: shared between multiple aggregators
\end{table*}

 \section[Communication protocols]{Communication protocols}
\label{sec:evci-review-protocols}

\begin{figure}[!ht]
  \begin{center}
    \includegraphics[width=\columnwidth]{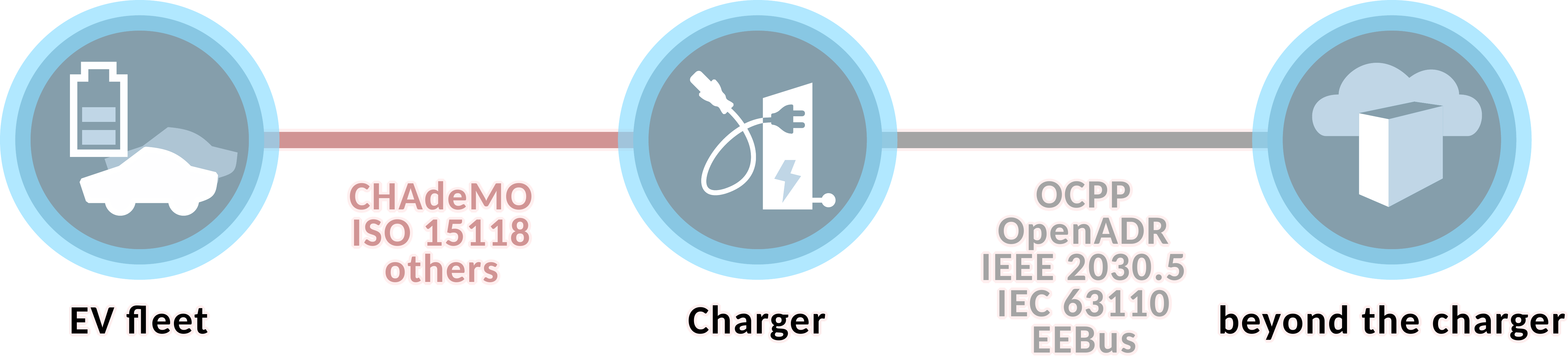}
  \end{center}
  \caption{Communications protocols in EV charging infrastructures.}
  \label{fig:ec-infrastructure-protocols}
\end{figure}

Protocols that have been proposed in EV charging infrastructures can be grouped into front-end and back-end protocols\cite{neaimeh2020mind}, as illustrated in Fig.~\ref{fig:ec-infrastructure-protocols}.
Front-end protocols specify the communication between the EVs and the chargers, while the back-end protocols specify the communication between the charger and other entities, or among other entities beyond the charger.
In this section we review some front-end protocols, i.e.,  ISO 15118 and  CHAdeMO, and comment on proprietary protocols.
Afterward, we review back-end protocols, i.e., OCCP, IEC 63110, OpenADR, IEEE 2030.5, and EEBus.
Finally, we compare these protocols and draw some security considerations.

\subsection{Security analysis of communication protocols}
\label{sec:security-analysis-communication-protocols}
Before discussing in details the security of protocols in the EV ecosystem, it might be useful to recall how security analysis of communication protocols is currently done.
In fact, we do not attempt to discuss security challenges of protocols from a technical point of view, but limit ourselves to provide an overview of their current configuration.
Before that, we notice that the most commonly used suite of (cryptographic) protocols to secure communication in the EV charging infrastructure is TLS, as illustrated in Section~\ref{sec:summary-protocols}.
However, we remark that {\em using TLS} is not enough, security can be jeopardised by either an improper configuration of TLS itself or its composition with other (even secure) protocols.
To appreciate the mechanics of compositional security flaws and the boundary of a security property, we provide an intuitive example.
The popular WhatsApp provides end-to-end secure communications guaranteeing the confidentiality of the messages; however, as soon as you enable backups to third parties, e.g. Google Drive, then all ``messages you back up are not protected by WhatsApp end-to-end encryption while in Google Drive''.
Therefore, the backup exposes the unencrypted messages to another communication protocol (composition) that exports them to somewhere else, and whose security cannot be guaranteed by the previously used protocol.

Commonly, the first security analysis of a communication protocol is proposed by their authors as a mathematical discussion over some desirable properties (e.g., secrecy, authentication) that depend on the context (e.g., cloud storage, automotive, mobile application) and the purpose or application of the protocol (e.g., management of charging schedule).
Security of protocols in the EV charging infrastructure is not an exception, and some of the most used protocols do not even attempt to cope with (information) security\cite{chademo}.

State-of-the-art analysis of cryptographic protocols (in this and other contexts) currently provides a mechanised mathematical model of the specifications along with a reproducible, automatic analysis of (some) security properties\cite{bhargavan2017verified,cremers2017comprehensive,metere2017automated,basin2018formal,hao2018analyzing}.
The related tools make use of formal languages and are familiar only to some security experts, who may still find it difficult to choose the right tools for the protocol to analyse (a useful comparison and guidelines are from Barbosa et al.\cite{barbosa2021sok}).
Lauser et al.\cite{lauser2020security} appreciated the importance of using the most popular formal security tools for the protocols used in the automotive industry; but a security expert is still needed to use such tools properly.
To enable non experts to approach verification of security or protocols, the modelling task needs to be automated too.
Unfortunately, the literature shows little effort toward automating the modelling task\cite{perrin2018noise,arnaboldi2019poster,kobeissi2020verifpal,metere2022automating}.
Still, their methodology either is not mature enough to be adopted by non experts or it is not ready to be used outside of the academic community.

As a final remark, we stress that the formal verification process of security protocols, when possible, is very difficult and time consuming.
Thus, it is common that researchers tend to focus only on the most {\em popular} protocols, where their research could be impactful.

\subsection{Front-end Protocols: EV to Charger}
\label{sec:front-end-protocols}

\subsubsection{CHAdeMO}
One of the first charging protocols offering bidirectional energy transfer for EVs is CHAdeMO.
It works in direct current and uses the internal communication network of modern cars, the CAN bus.
While CHAdeMO implements safety, it does not implement secure communications.
CHAdeMO follows the standard IEC 61851-23:2014 for the actual charging and discharging operations, and follows the standard IEC 61851-24:2014 for the digital communication between the car and the charger.
When an EV plugs in with the CHAdeMO cable, it connects directly its CAN bus to the (untrusted) charger, and messages are exchanged unencrypted.
Hence, security breaches of the CAN bus directly affect the CHAdeMO protocol.
We refer to Metere et al.\cite{metere2021securing} for a description of its vulnerabilities.

\subsubsection{ISO 15118}
A more recent front-end protocol is ISO 15118.
ISO 15118 can, but is not limited to, work over the Combined Charging System (CCS), which is a set of hardware and software standards for charging systems.
One main characteristic of ISO 15118 is the specification of digital certificates to authenticate devices.
With them it offers seamless services, as Plug\&Charge (PnC), allowing for automated authentication and authorization.
To offer security, ISO 15118 {\em suggests}\footnote{"TLS is not mandatory for certain Identification Modes other than the Plug-and-Charge Identification Mode", from ISO15118-2:2014 (revised and reconfirmed in 2020).} the adoption of TLS.
Eavesdropping wireless attacks on the physical-layer of the CCS\cite{baker2019losing} have been demonstrated; such attacks can be avoided by mandating the use TLS.

\subsubsection{Security of Proprietary protocols: the example of Tesla}
Some companies, e.g., Tesla, implemented their own alternative front-end protocols.
An analysis based on reverse engineering\cite{mahaffey2015hacking} showed that their protocol works over the CAN bus.
The general architecture of the CAN bus shows a central gateway that connect other buses with electronic units, e.g., ABS, and telematic units, e.g., Bluetooth and WiFi modules.
In a work from Nie et al.\cite{nie2017free}, authors describe how they exploited an anomaly in the gateway that allowed them to use wireless communication to inject malicious messages into the CAN bus.
The attack was so devastating that they could gain remote control of the car.
Interestingly, this vulnerability was patched by Tesla in only 10 days.

\subsection{Back-end Protocols: Charger to third party operators}
\label{sec:back-end-protocols}

\subsubsection{OCPP}
The Open Charge Point Protocol (OCPP) provides accessibility, compliance, and uniform communications between electric vehicle charging stations (CS) and charging station management systems (CSMS), with no cost or licensing barriers.
This protocol is becoming the de facto standard protocol and enables customers to switch the charging network without replacing the charging station.
As of 2019 the use of OCPP 1.6 (or equivalent) is required for new charging stations in the UK.
Since this version, an Internet connection is enabled and potentially increases the devices' attack surface.
The latest version, OCPP 2, improves the security measures: secure connection setup, security events/logging, and secure firmware update.
Only recently, the open charge alliance published a standard way to address (some) security using OCPP 1.6-J, as before then the security design was fully delegated to individual implementers of OCPP (in other words, no security was specified).
OCPP 2 can operate according to three security profiles, illustrated in Table~\ref{tbl:ocpp-security-profiles}: Unsecured Transport with Basic Authentication, TLS with Basic Authentication and TLS with Client-Side Certificates.
\begin{table}[htbp]
  \caption{OCPP security profiles.}
  \label{tbl:ocpp-security-profiles}
  \renewcommand{\rotation}{90}
  \small\sffamily
  \centering
  \renewcommand{\arraystretch}{1.1}
  \setlength\arrayrulewidth{2pt}
  \arrayrulecolor{white}
  \rowcolors{2}{evenrows}{evenrows}
  \begin{tabular}{l|c|c|c}
   &
  \multicolumn{2}{c|}{\thh{\multilinecell[headercolor]{c}{l}{Authentication}}} &
  \cellcolor{headercolor} \\
  
  \thh{\rowcolors{2}{headercolor}{headercolor}
    \renewcommand{\arraystretch}{1.1}
    \begin{tabular}{l}
      Profile
    \end{tabular}} &
  \thh{\multilinecell[headercolor]{b}{c}{Charging\\Station}} &
  \thh{\multilinecell[headercolor]{b}{c}{Charging\\Station\\Management\\System}} &
  \cellcolor{headercolor}\thr{Security} \\
  \hline

  \thh{\multilinecell[headercolor]{c}{l}{Unsecured\\Transport\\with Basic\\Authentication}}
    & credentials
    & trusted
    & -- \\ \hline
  \thh{\multilinecell[headercolor]{c}{l}{TLS with Basic\\ Authentication}}
    & credentials
    & \multilinecell{c}{c}{X509\\certificate}
    & TLS \\ \hline
  \thh{\multilinecell[headercolor]{c}{l}{TLS with\\Client-Side\\ Authentication}}
    & \multilinecell{c}{c}{X509\\certificate}
    & \multilinecell{c}{c}{X509\\certificate}
    & TLS \\ \hline
  \end{tabular}
\end{table}

\subsubsection{IEEE 2030.5}
IEEE 2030.5 is a standard whose design is dedicated and optimised for devices in the same home area network.
In relation with the EV charging infrastructure, it specifies communication protocol suitable for being used between most entities: aggregators, home-smart devices, chargers, EVs.

Security is a first-class citizen in IEEE 2030.5, differently from some of its predecessors, e.g., IEC 61850.
In IEEE 2030.5 devices embed a life-long certificate that cannot be renewed or revoked and must be exchanged privately.
This is in strong contrast with a PKI that is thus not supported by the standard.
We remark that revocation is a crucial feature.
An example from the Internet, in 2014 a bug affecting the TLS protocols was discovered and affected a significant number of server and websites.
As a consequence, the authenticity of their certificate could not be trustable any longer and needed to be revoked and reissued\cite{zhang2014analysis}.
For this reason, IEEE 2030.5 does not currently scale and therefore is not applicable to the whole EV charging infrastructure.

\subsubsection{OpenADR}
At a higher level if compare with OCPP, we have Open Automated Demand Response (OpenADR).
It is an information exchange model for distributed energy resources (DERs).
Still on a high level, it generally relies on a (centralised) gateway, or an aggregator to translate utility Demand Responses (DR) and DER requirements into specific device behaviours.

Security of OpenADR is provided by the (mandatory) use of TLS.
TLS is used to offer integrity (including authentication) and confidentiality.
Legacy devices might not be compliant with OpenADR due to limited resources, e.g., insufficient processing power or network bandwidth.
Through TLS, an implementation of the PKI for OpenADR (by Kyrio, Inc. a subsidiary of CableLabs) creates an ecosystem to trust identities through digital certificates\footnote{More details on PKIs are provided in Metere et al.\cite{metere2021securing}.}.

In OpenADR, the architecture of entities is abstracted (virtual nodes), taking an indirect approach toward device control, in contrast with the approach of IEEE 2030.5.
Some of the functionality offered by OpenADR and IEEE 2030.5 overlap; though, their interoperability is possible and (just) requires some isolation between the two, e.g., avoiding controlling the same devices with incompatible decisions.

\subsubsection{IEC 63110}
IEC 63110 is a back-end protocol dedicated to charging and  discharging operations of EVs, still under development~\footnote{\url{https://www.ncl.ac.uk/media/wwwnclacuk/cesi/files/20200115_Meet IEC 63110_ Paul Bertand SmartFuture-min.pdf}}.
Their approach to cyber security is direct, as they mandate the use of TLS for all communications; in particular, it offers end-to-end security between clients and servers.
Additionally, IEC63110 can natively interoperate with ISO 15118 (with TLS) and fully supports its PKI.

\subsubsection{EEBus}
Due to some similarities that the EV charging infrastructure has with communications of Internet of Things (IoT) devices, another suite of protocols that could be possibly be used is EEBus.
EEBus emphasises on what data structure is used for communication, and how they are exchanged among entities, and many companies adopted it.
The suite of protocols specified in EEBus work at different levels: the SPINE (Smart Premises Interoperable Neutral-message Exchange) protocol works at the information layer and the SHIP (Smart Home IP) protocol works at the communication layer.
SHIP implements TLS to provide a secure TCP/IP-based solution.
To provide security, SPINE runs over SHIP.
Overall, EEBus can operate along with the Internet and can be, in principle, used for securing communications in the EV charging infrastructure.
Even if EEBus specifications are free and open, no existing actual implementations is freely available\footnote{\url{https://medium.com/grandcentrix/will-microsoft-joining-eebus-finally-bring-us-an-open-source-reference-implementation-3432a93dd2e6}}.
 \subsection{Summary of protocols in the EV ecosystem}
\label{sec:summary-protocols}
The Table~\ref{tbl:summary-protocols} summarises front-end and back-end protocols reviewed in Sections~\ref{sec:front-end-protocols} and~\ref{sec:back-end-protocols}.
\begin{table}[!ht]
  \caption{Summary of the security of communication protocols in EV charging infrastructures. \\Legend: $\rc$~--~none; $\rh$~--~partial or optional; $\rt$~--~partial (in security, TLS without no PKI); $\rb$~--~full (in security, TLS and PKI).}
  \label{tbl:summary-protocols}
  \renewcommand{\rotation}{90}
  \small\sffamily
  \centering
  \renewcommand{\arraystretch}{1.2}
  \setlength\arrayrulewidth{2pt}
  \arrayrulecolor{white}
  \rowcolors{2}{evenrows}{evenrows}
  \begin{tabular}{ccc|c|c|c|l}
  \clearcell &
  \multicolumn{2}{c|}{\thh{\cellcolor{evenrows}Protocol        }} &
                      \thr{\cellcolor{evenrows}Security        } &
                      \thr{\cellcolor{evenrows}Open specs      \ \ } &
                      \thr{\cellcolor{evenrows}Open code       \ \ } &
                      \thh{\cellcolor{evenrows}Interoperability} \\
  \hline

  \clearcell
  & \multicolumn{2}{c|}{\thh[headercoloralt]{\multilinecell[headercoloralt]{c}{c}{CHAdeMO}}}
    & $\rc$
    & $\rh$
    & $\rh$
    & -- \\ \hline
  
  \multirow{-5}{*}{\clearcell\thr[90]{Front-end\quad}}
  & \multicolumn{2}{c|}{\thh[headercoloralt]{\multilinecell[headercoloralt]{c}{c}{ISO 15118}}}
& $\rh$
    & $\rb$
    & $\rb$
    & -- \\
  \arrayrulecolor{headercoloralt}
  \hline
  \arrayrulecolor{white}
  \hline
  
  \clearcell
  & \thh{}
  & \thh{2.0}
& $\rb$
  & 
  & 
& 
  \\

  \clearcell
  & \thh{OCPP}
  & \thh{1.6-J}
& $\rh$
  & $\rb$
  & $\rb$
& \cellwithnote{$\rb$}{ISO 15118 TLS}
  \\

  \clearcell
  & \thh{}
  & \thh{$\leq$ 1.6}
& $\rc$
  & 
  & 
& 
  \\ \hline
  
  \clearcell
  & \multicolumn{2}{c|}{\thh{\multilinecell[headercolor]{c}{c}{IEC 63110}}}
& $\rb$
    & $\rb$
    & --
    & \cellwithnote{$\rb$}{ISO 15118 TLS}
  \\ \hline
  
  \clearcell
  & \multicolumn{2}{c|}{\thh{\multilinecell[headercolor]{c}{c}{OpenADR}}}
& $\rb$
    & $\rb$
    & $\rb$
& \cellwithnote{$\rb$}{IEEE 2030.5}
  \\ \hline
  
  \clearcell
  & \multicolumn{2}{c|}{\thh{\multilinecell[headercolor]{c}{c}{IEEE 2030.5}}}
& $\rt$
    & $\rb$
    & $\rb$
& \cellwithnote{$\rb$}{OpenADR}
  \\ \hline
  
  \multirow{-4}{*}{\clearcell\thr[90]{Back-end}}
  & \multicolumn{2}{c|}{\thh{\multilinecell[headercolor]{c}{c}{EEBus}}}
& $\rb$
    & $\rb$
    & $\rc$
    & $\rh$
  \\
  \arrayrulecolor{headercolor}
  \hline
  \end{tabular}
\end{table}
The table shows the status of each protocol in terms of cyber security, openness, and interoperability.
This summary reflects how cyber security has been considered in the communication network of EV charging infrastructure.
Although the focus of this work is on cyber security, we also provide status about open specs, open code, and interoperability to offer a comprehensive reference for future practice by researchers in this area.
 \section{Related work}
\label{sec:related-work}
Several studies have investigated cyber security and privacy challenges and their impact on interconnected energy systems.
This paper significantly extends beyond the work of Pourmirza and Walker\cite{pourmirza2021electric}
and, partly, Metere et al.\cite{metere2021securing}.
On top of them, we provide a single comprehensive and detailed overview of security and privacy on EV charging infrastructure.
We propose a detailed review of front-end and back-end protocols currently used in the EV charging infrastructure.
Beyond that, we mention additional attack types and examples.
We also extend the discussion on standards for EV infrastructure by including PAS 1878 and PAS 1879, to which the authors contributed.

On the same topic as this work, Khalid et al.\cite{khalid2019facts} showed their concern on a lack of standards addressing cyber security challenges.
However, their research restricts to battery energy management only, while we cover the whole EV ecosystem.
A relatively less recent work on the same lines\cite{mustafa2013smart} analyses security limiting to a brief discussion of potential threats on a proposed design of the EV ecosystem.
After a decade of advancements in the field, we can refer to protocols used in the real world.

An overview of cyber security issues related to EVs has been discussed by Fraiji et al.\cite{fraiji2018cyber}, where they highlighted the challenges of the Internet of Electric Vehicles, i.e., vehicles, sensors, humans, road infrastructure, and charging stations.
Their discussion does not include the Energy sector.
With close relation to the EV charging infrastructure, Reeh et al.\cite{reeh2019vulnerability} assessed vulnerability and risks of cyber-physical attacks that can potentially affect the specific EV charging infrastructure of UCLA WinSmartEV\texttrademark.
Though, their study is far from being comprehensive as their analysis is based on a collection of potential attacks in similar contexts, without taking into consideration {\em any} of the protocols that are commonly used in the EV charging infrastructure.

Probably the closest (and most recent) works to ours are by Antoun et al.\cite{antoun2020detailed} and by Van Aubel and Poll\cite{vanaubel2022security}.
They survey and link attacks and proposed mitigations found in the literature on the EV charging ecosystem.
However, they focus on the OCPP protocol (EV charger management) and do not review other back-end protocols used in the EV charging infrastructure, which are reviewed in this paper.
Van Aubel and Poll\cite{vanaubel2022security} share with us the goal of providing a cyber security analysis; however, ours is more comprehensive of protocols and we also add privacy concerns, see Section~\ref{sec:evci-privacy}.
Our work relates to the situation of EV charging in the UK, their work could be (partly) used to compare with the situation of charging infrastructure Netherlands.

 \section{Conclusions and future directions}
\label{sec:conclusion}

In this paper we explored cyber security of EV charging infrastructure by providing an overview of system vulnerabilities, key challenges, and its requirements.
This work is the most comprehensive study to date of cyber security for EV charging infrastructure.
Moreover, we discussed privacy concerns for the users and data required to be exchanged by various entities.
Finally, we provided a comprehensive review of front-end and back-end communication protocols that are currently used, or are in active development. 

Overall, we recommend considerations of the whole system approach rather than focusing on one particular entity.
The whole system should cover people (consumers), physical environment (settings where EV ecosystem components are installed including consumer premises, service provider and cloud/edge environment), process (business process used to deliver services and control decisions), and technology (technical issues for design and development of secure components and data).
We can appreciate a tendency to adopt security by design, as in recent regulations proposed in similar contexts, e.g., smart meters or IoT.

Future research on the security related to data and communications in the EV charging infrastructure can be carried out with existing tools and techniques.
Not surprisingly, such tools are already used to analyse or evaluate the security of the larger smart grid or a portion of it, e.g., cyber-physical systems.
We suggest the following reading to explore similar tools~\cite{czekster2021systematic}.
Apart from those mentioned in Section~\ref{sec:security-analysis-communication-protocols} that are suitable for technical analysis of specific protocols, a commonly used methodology is simulation of the power network or the communication network, or a co-simulation of both of them.
Recent efforts~\cite{morstyn2020open,nasr2022power} suggest that co-simulation is suitable for modelling security flaws and, in particular, their effects to the voltage, current or frequency in the simulated system, that can be devastating~\cite{lee2016analysis,soltan2018blackiot}.
As in other realities of the smart grid~\cite{czekster2022incorporating}, cyber-threat intelligence would be desirably collected by security practitioners, solutions as intrusion detection/prevention systems would be adopted, and proper physical security would be implemented.

\section*{Acknowledgment}

This research was partly funded by The Alan Turing Institute Data-Centric Engineering Programme under a grant from the Lloyd's Register Foundation (G0095), an Innovate UK e4Future grant (104227), under a grant from EPSRC for Active Building Centre (EP/V012053/1), and supported by the Industrial Strategy Challenge Fund under the programme grant EP/S016627/1.
The authors would like to thank Charles Morisset, Carsten Maple, Xavier Bellekens, Ricardo M. Czekster for their review and comments that made part of our narrative more fluid and rigorous.

\ifCLASSOPTIONcaptionsoff
  \newpage
\fi

\bibliographystyle{IEEEtran}
\bibliography{references}

\end{document}